\documentclass[12pt]{article}

\usepackage{soul}
\usepackage{graphicx} 
\usepackage{gensymb}

\usepackage{amssymb,amsmath,bm,txfonts}
\usepackage{mathrsfs} 
\usepackage{siunitx} 
\usepackage{textcomp}
\usepackage{xcolor} 
\usepackage{chemformula} 
\graphicspath{ {images/} }

\usepackage{scicite}

\usepackage{times}

\newenvironment{sciabstract}{%
\begin{quote} \bf}
{\end{quote}}

\title{Elastic turbulence generates \\anomalous flow resistance in porous media}

\author
{Christopher A. Browne$^{1}$ and Sujit S. Datta$^{1\ast}$
\\
\normalsize{$^{1}$Department of Chemical and Biological Engineering, Princeton University}\\
\normalsize{41 Olden Street, Princeton, NJ 08544, USA}
\\
\normalsize{$^\ast$To whom correspondence should be addressed; E-mail:  ssdatta@princeton.edu.}
}

\date{}

\begin{document} 


\baselineskip24pt


\maketitle

\begin{sciabstract} 

Diverse processes rely on the viscous flow of polymer solutions through porous media. In many cases, the macroscopic flow resistance abruptly increases above a threshold flow rate in a porous medium---but not in bulk solution. The reason why has been a puzzle for over half a century. Here, by directly visualizing the flow in a transparent 3D porous medium, we demonstrate that this anomalous increase is due to the onset of an elastic instability. We establish that the energy dissipated by the unstable flow fluctuations, which vary across pores, generates the anomalous increase in flow resistance through the entire medium. Thus, by linking the pore-scale onset of unstable flow to macroscopic transport, our work provides generally-applicable guidelines for predicting and controlling polymer solution flows.\\

One-sentence summary: Microscopy reveals that chaotic fluctuations control the flow resistance of a polymer solution in a porous medium.\\

\end{sciabstract}

    \newpage\noindent Diverse applications, ranging from groundwater remediation \cite{smith2008} and oil recovery \cite{sorbie2013,durst1981} to filtration \cite{bourgeat2003filtration} and chromatography \cite{luo1996high}, rely on the viscous-dominated flow of polymer solutions through disordered porous media. One of the most fundamental descriptors of such flows is the ``apparent viscosity" $\eta_{\text{app}}$, which quantifies the macroscopic resistance to flow through the tortuous pore space. At low flow rates, $\eta_{\text{app}}\approx\eta$, the dynamic shear viscosity of the bulk solution. Above a threshold flow rate, however, $\eta_{\text{app}}$ abruptly \textit{increases} for many polymer solutions---even though the shear viscosity $\eta$ of the bulk solution \textit{decreases} with increasing shear rate \cite{james1975laminar,durst1981,clarke2016}. The reason for this anomalous increase has remained a puzzle ever since it was first reported over half a century ago \cite{marshall1967flow}.  While many mechanisms have been proposed \cite{browne2019pore}, assessing their influence has been challenging; typical 3D media are opaque, precluding direct characterization of the flow \textit{in situ}. As a result, despite its fundamental importance and strong impact in applications, why the macroscopic flow resistance of polymer solutions anomalously increases in porous media is still unknown. Here, by directly visualizing the flow of a polymer solution in a transparent 3D porous medium, we demonstrate that this anomalous increase is due to the added dissipation arising from an elastic instability.

Our porous medium is a consolidated random packing of borosilicate glass beads (Figure 1A). The fluid used is a dilute solution of 18 MDa partially hydrolyzed polyacrylamide in a viscous aqueous solvent, formulated to precisely match its refractive index to that of the glass beads---rendering the medium transparent when saturated. Additionally dispersing a dilute fraction of $200$ nm diameter fluorescent latex microparticles, which act as flow tracers \cite{SI}, therefore enables measurement of the two-dimensional (2D) fluid velocities $\textbf{u}$ in the pore space \textit{via} particle image velocimetry (PIV) using confocal microscopy. We characterize the macroscopic flow behavior by injecting the polymer solution into the medium at a constant volumetric flow rate $Q$ and measuring the corresponding steady-state pressure drop $\langle\Delta P\rangle_t$ across the medium; the subscript indicates an average over time $t$. For a Newtonian fluid, the relationship between these quantities is given by Darcy's law: $\langle\Delta P\rangle_{t}/\Delta L =\eta (Q/A)/k$, where $\eta$ is the fluid dynamic shear viscosity and $\Delta L$ and $k$ are the length and absolute permeability of the medium, respectively. For a polymer solution, for which the viscosity can change depending on flow conditions, this relationship is still employed in practice, but with $\eta$ replaced by the ``apparent viscosity" $\eta_{\text{app}}\equiv\frac{\langle\Delta P\rangle_{t}/\Delta L}{ (Q/A)/k}$ that represents the macroscopic flow resistance. To facilitate comparison to the bulk shear viscosity, we therefore represent the pressure drop measurements by plotting the reduced apparent viscosity $\eta_{\text{app}}/\eta(\dot{\gamma}_I)$ as a function of the interstitial shear rate $\dot{\gamma}_I\equiv (Q/A)/\sqrt{\phi k}$ defined using the characteristic pore length scale $\sqrt{\phi k}$, where $\phi$ is the porosity of the medium \cite{zami2016transition,berg2017shear}. As expected, at low flow rates, $\eta_{\text{app}}=\eta(\dot{\gamma}_I)$. However, above a critical flow rate corresponding to $\dot{\gamma}_{I}\approx5$ $\mathrm{s}^{-1}$, $\eta_{\text{app}}$  increasingly exceeds $\eta(\dot{\gamma}_I)$, eventually peaking at $\approx5\eta(\dot{\gamma}_I)$ as shown in Fig. 1B. This anomalous increase in the macroscopic flow resistance parallels previous reports \cite{james1975laminar,durst1981,clarke2016,marshall1967flow}.

\begin{figure}
    \centering
    \includegraphics[width=\textwidth]{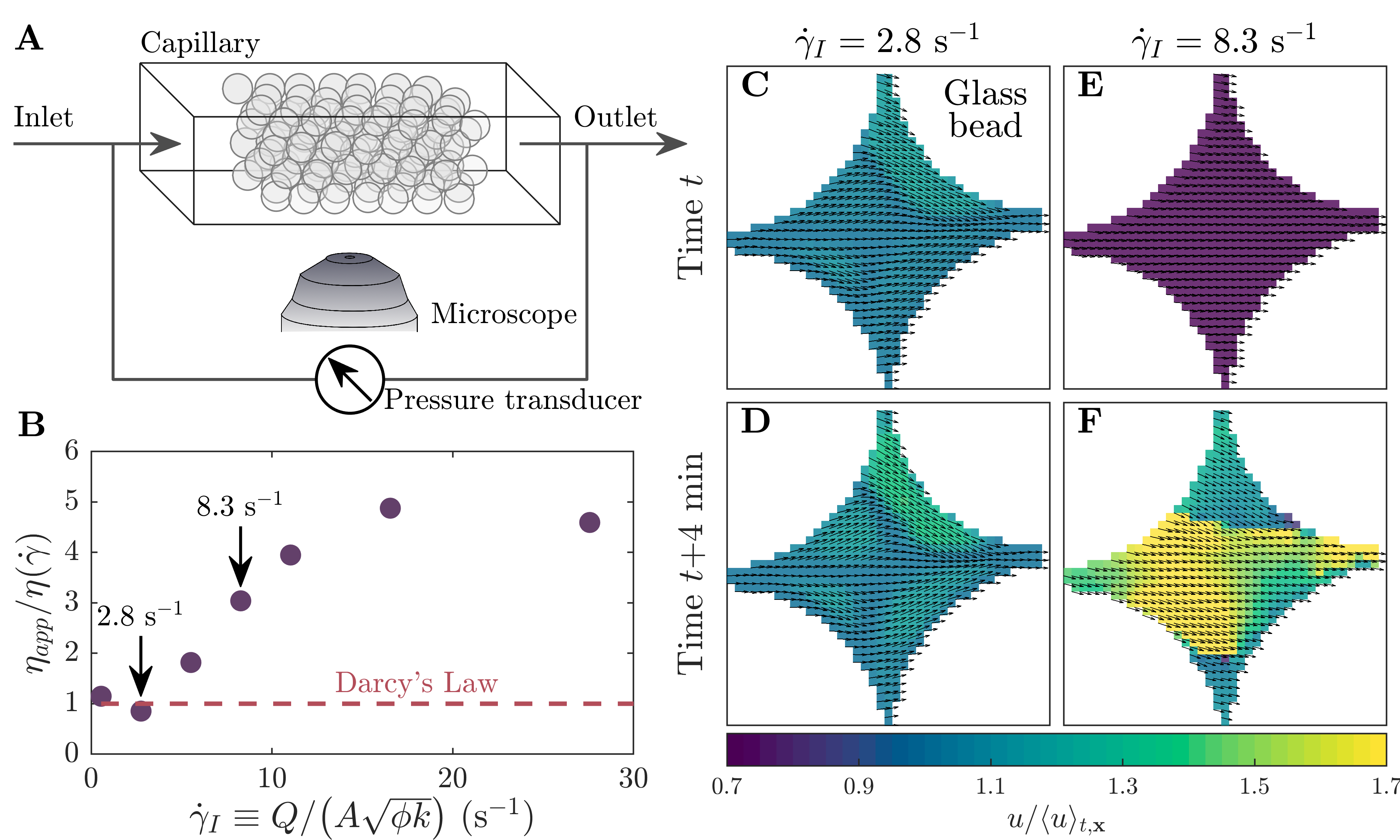}    
    \caption{\textbf{Pore-scale visualization reveals that the anomalous increase in flow resistance coincides with the onset of elastic turbulence}. \textbf{(A)} Our porous medium is a lightly-sintered random packing of borosilicate glass beads $D_p=330\pm30~\muup$m in diameter, with porosity $\phi\approx0.41$ and length $\Delta L=8.1$ cm. The packing is confined in a quartz capillary with square cross section $A=3.1~\mathrm{mm}\times3.1~\mathrm{mm}$. The fluid used is a dilute solution of 18 MDa partially hydrolyzed polyacrylamide in 81 wt\% glycerol, 12 wt\% DMSO, 6 wt\% deionized water, and 1 wt\% NaCl. We inject the solution into the medium using a syringe pump, and simultaneously image the flow \textit{in situ} using a confocal microscope while measuring the pressure drop across the medium using differential pressure transducers. \textbf{(B)} Above a threshold flow rate, parameterized by the characteristic shear rate $\dot{\gamma}_I$, the macroscopic pressure drop anomalously increases and deviates from the prediction of Darcy’s Law given by the shear viscosity of the bulk solution. \textbf{(C-F)} Flow visualization in an example pore; applied flow is left to right. Arrows indicate the vector field, and colors indicate velocity magnitude. \textbf{(C-D)} At a low flow rate, the flow does not change over time. \textbf{(E-F)} At a higher flow rate, the flow exhibits strong spatio-temporal fluctuations.}
    \label{fig:Setup}
\end{figure}

Simultaneous visualization of the pore-scale flow hints at the underlying reason for this anomalous increase. Fig. 1C--F shows the velocity field within an example pore measured at two different times. At low flow rates, for which $\eta_{\text{app}}=\eta(\dot{\gamma}_I)$, the flow is laminar and steady over time (Fig. 1C--D, Movie S1). Strikingly, concomitant with the anomalous increase in flow resistance, we observe strong spatial and temporal fluctuations in the flow at high flow rates (Fig. 1E--F)---despite the negligible influence of inertia in the flow, as indicated by the Reynolds number Re $\lesssim10^{-4}\ll1$. As shown in Movie S2, the fluid pathlines continually cross and vary over time, indicating the emergence of an elastic instability. This instability---often known as ``elastic turbulence'' due to the similarity of the chaotic flow field to that of inertial turbulence \cite{groisman2000,pan2013,qin2019flow}---is generated by the buildup of elastic stresses as the polymers are elongated by the flow through the tortuous pore space. It is well-studied in a range of simplified geometries \cite{larson1990,shaqfeh1996,pakdel1996, rodd2007,afonso2010purely,zilz2012,galindo2012,pan2013,ribeiro2014,kawale2017a,sousa2018purely,qin2019flow,Browne2020,browne2019pore}. Our visualization reveals that it also arises in 3D porous media---contrary to previous predictions, based on studies in 2D media, claiming that the disordered structure of the medium suppresses elastic turbulence \cite{walkama2020disorder}.

To further characterize the conditions under which elastic turbulence arises in this pore, we subtract the mean from each velocity vector to focus on the fluctuations, $\textbf{u}'=\textbf{u}-\langle\textbf{u}\rangle_{t}$, where the subscript indicates an average over time $t$. Near the onset of the anomalous increase in flow resistance, flow fluctuations (blue in Fig. 2A-B) manifest as intermittent, abrupt bursts that coexist with the base laminar flow (purple in Fig. 2B), but quickly decay (Movie S3). Well above this onset, however, these fluctuations (blue-green-yellow in Fig. 2C) still coexist with the laminar flow, but persist over time (Fig. 2D, Movies S4--S5). Intriguingly, similar behavior is observed in the transition to inertial turbulence; near the transition, discrete bursts of unstable flow appear and decay, while above the transition at sufficiently large Re, these bursts percolate through time \cite{lemoult2016directed}. Thus, the transition to inertial turbulence is thought to be a non-equilibrium phase transition in the directed percolation (DP) universality class \cite{pomeau1986front}. Our results suggest the tantalizing possibility that the pore-scale transition to elastic turbulence may similarly be a non-equilibrium phase transition, as suggested recently in simulations \cite{van2018elastic}.

\begin{figure}
    \centering
    \includegraphics[width=.9\textwidth]{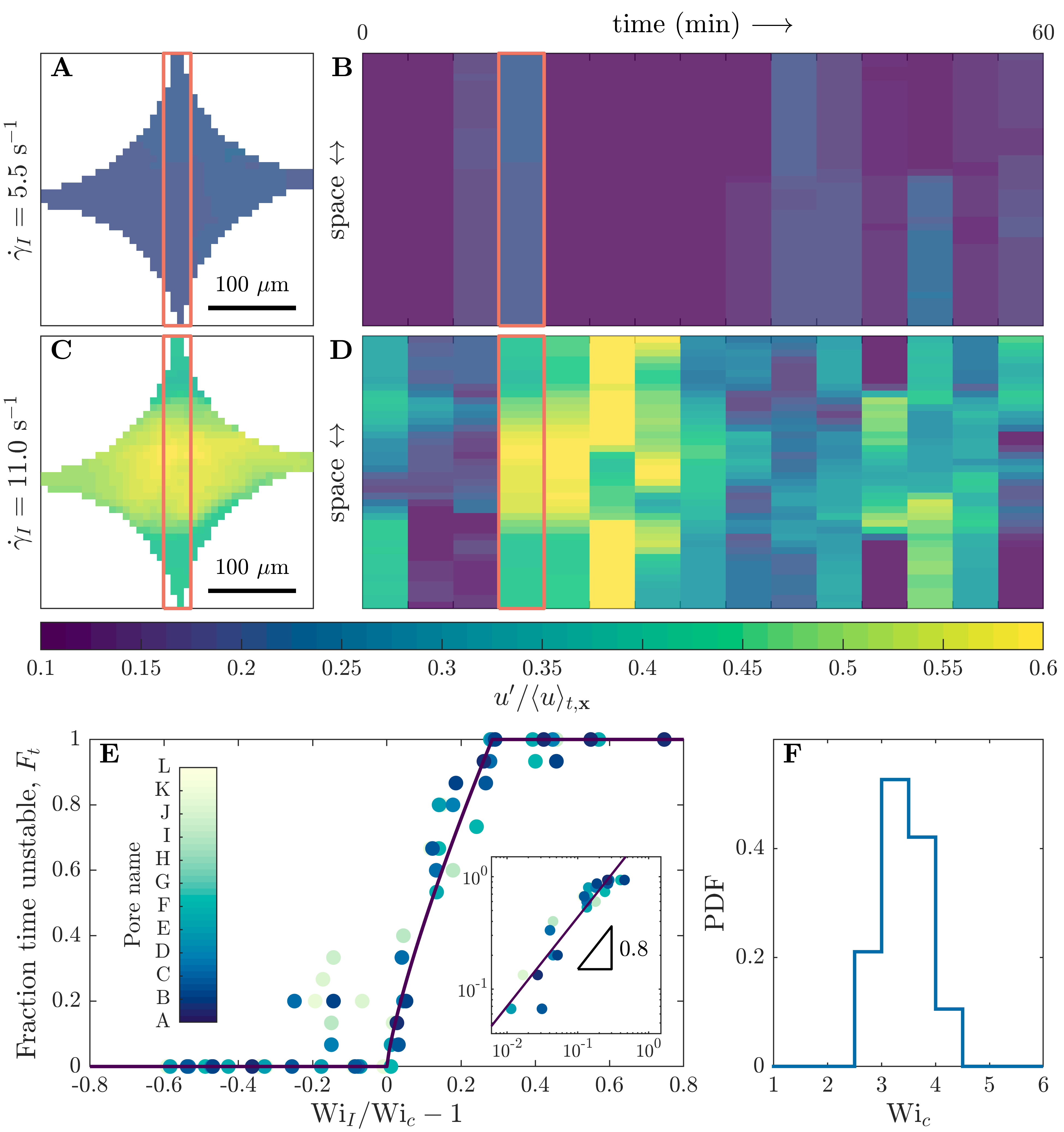}
    \caption{\textbf{The pore-scale transition to elastic turbulence is a non-equilibrium phase transition}. \textbf{(A-B)} Near the onset of elastic turbulence, flow fluctuations are intermittent and short-lived; \textbf{(A)} shows the magnitude of flow fluctuations in a given pore at a given time, while \textbf{(B)} shows how the fluctuations in the red box vary over time. \textbf{(C-D)} Well above the onset of elastic turbulence, flow fluctuations are stronger and persist over time. \textbf{(E)} The fraction of time $F_t$ a pore spends in an unstable state ($u'/\langle u\rangle_{t,\mathbf{x}}>0.2$) continually grows above a threshold flow rate, parameterized by the threshold Weissenberg number $\mathrm{Wi}_c$. Different pores are characterized by different values of $\mathrm{Wi}_c$, as shown by the probability density function in \textbf{(F)}; however, they all exhibit a similar transition to elastic turbulence as shown by the collapse of the measurements of $F_t$ in \textbf{(E)} when the imposed $\mathrm{Wi}_I$ is rescaled by $\mathrm{Wi}_c$ for each pore. The inset shows the power-law scaling $F_t\sim\left(\mathrm{Wi}_I/\mathrm{Wi}_c-1\right)^{0.8}$.}
    \label{fig:Kym}
\end{figure}

We test this hypothesis by measuring the fraction of time $F_{t}$ a pore spends in the unstable state, as is done to characterize the transition to inertial turbulence \cite{barkleyjfm}, for 12 different pores over a broad range of flow rates.  Because the unstable flow is driven by polymer elasticity rather than fluid inertia, we describe the pore-scale transition to elastic turbulence using the characteristic Weissenberg number defined using the macroscopic imposed flow conditions, $\mathrm{Wi}_I\equiv N_1(\dot{\gamma}_I)/2\sigma(\dot{\gamma}_I)$; this parameter compares elastic stresses quantified by the first normal stress difference $N_{1}$ to viscous stresses quantified by the shear stress $\sigma$, and represents the upper limit of the spatially-varying local Weissenberg number \cite{SI}. It can also be related to the largest destabilizing term in a linear stability analysis of the Stokes equation for a viscoelastic fluid \cite{SI,pakdel1996}. For each pore, at low $\mathrm{Wi}_I$, the flow is laminar and unchanging in time, with $F_{t}=0$. Above a critical value $\mathrm{Wi}_{c}$, however, the pore is unstable for a non-zero fraction of time, and $F_{t}$ smoothly increases above zero. It eventually saturates at unity for $\mathrm{Wi}_I\gg \mathrm{Wi}_{c}$, indicating that the elastic turbulence has fully developed. Notably, this transition is general: while the critical value $\mathrm{Wi}_{c}$ varies from pore to pore (Fig. 2E), presumably due to the disordered structure of the pore space, $F_{t}$ grows similarly with the rescaled $\mathrm{Wi}_{I}/\mathrm{Wi}_{c}$ for all 12 pores, as shown by the different colors in Fig. 2F---indicating a second-order phase transition. Furthermore, while there is some scatter, the data for $\mathrm{Wi}_{I}$ near $\mathrm{Wi}_{c}$ are consistent with the power-law scaling $F_{t}\sim\left(\mathrm{Wi}_{I}/\mathrm{Wi}_{c}-1\right)^{0.8}$ (Fig. 2F, inset), in good agreement with the scaling exponent of DP in 3 spatial dimensions, 0.81 \cite{dp}. Thus, the pore-scale transition to elastic turbulence is a non-equilibrium phase transition that appears to also fall in the DP universality class, similar to the case of inertial turbulence \cite{lemoult2016directed,van2018elastic}.

\begin{figure}
    \centering
    \includegraphics[width=\textwidth]{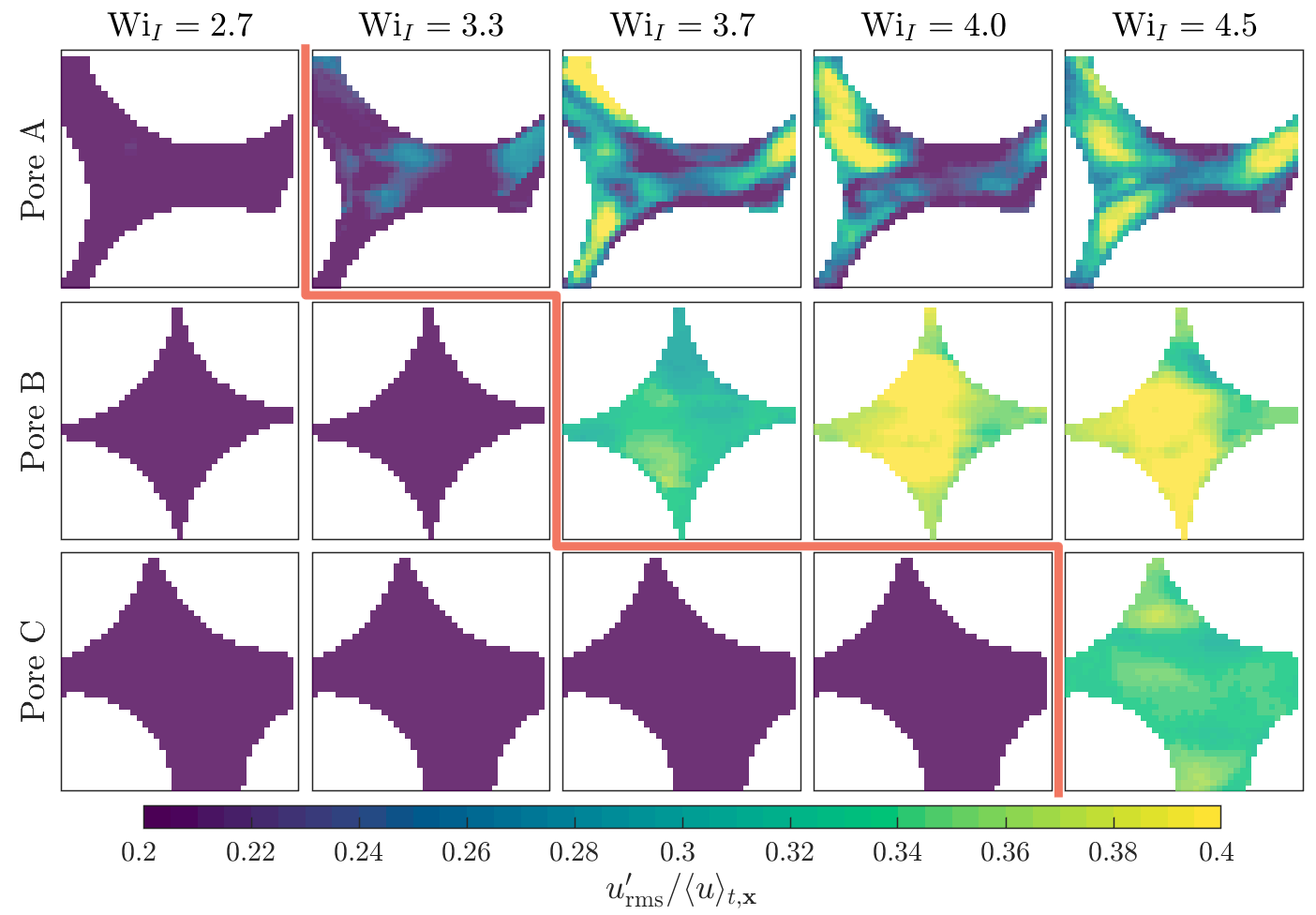}
    \caption{\textbf{The occurrence of elastic turbulence is spatially heterogeneous throughout a porous medium}. Images show the magnitude of the normalized root mean square flow fluctuations over 60 min in different pores and at different flow rates, parameterized by $\mathrm{Wi}_I$. Applied flow is left to right. Pore A becomes unstable at the lowest flow rate, as shown by the red line in the first row. Pore B becomes unstable at the next highest flow rate, shown by the red line in the second row. Pore C becomes unstable only at even higher flow rates. }
    \label{fig:Patches}
\end{figure}

An unexpected consequence of the pore-to-pore variability in $\mathrm{Wi}_{c}$, which ranges from $\mathrm{Wi}_{c,min}\approx 2.7$ to $\mathrm{Wi}_{c,max}\approx4.5$ (Fig. 2E), is that the occurrence of elastic turbulence is spatially heterogeneous throughout the medium. In particular, because some pores become unstable at different values of $\mathrm{Wi}_{I}$ than others, unstable pores coexist amid stable, laminar pores for $\mathrm{Wi}_{I}$ in this range. An example is shown in Fig. 3, which displays the normalized root mean square of the flow fluctuations $u'$ over time, $u'_{rms}/\langle u\rangle_{t,\mathbf{x}}$ for three different pores. At low flow rates and therefore $\mathrm{Wi}_{I}$, all three pores are stable, as shown in the first column. As flow rate and therefore $\mathrm{Wi}_{I}$ is increased, Pore A ($2.7<\mathrm{Wi}_{c}\leq3.3$) becomes unstable first, while Pores B--C remain stable, as shown in the second column. At a higher $\mathrm{Wi}_{I}$, Pore B ($3.3<\mathrm{Wi}_{c}\leq3.7$) next becomes unstable, as shown in the third column. Finally, at an even higher $\mathrm{Wi}_{I}$, Pore C ($4.0<\mathrm{Wi}_{c}\leq4.5$) also becomes unstable, as shown in the last column. Thus, as $\mathrm{Wi}_{I}$ increases from $\approx2.7$ to $4.5$, an increasing fraction of pores becomes unstable.

How does this variability in the occurrence of elastic turbulence impact the macroscopic flow resistance? Motivated by the similarities between elastic and inertial turbulence revealed by our pore-scale imaging, as well as by previous studies in a range of simplified geometries \cite{groisman2000,pan2013,qin2019flow}, we hypothesize that the flow fluctuations that arise in elastic turbulence impart additional viscous dissipation to the flow---akin to fluctuations in inertial turbulence. We quantify this hypothesis using the power density balance for viscous-dominated flow \cite{SI},
\begin{equation}
    -\nabla\cdot P\mathbf{u}=\boldsymbol{\tau}:\nabla\mathbf{u},
    \label{eq1}
\end{equation}

\noindent where the left hand side represents the rate of work done by the fluid pressure and the right hand side represents the rate of viscous energy dissipation, per unit volume; here $\boldsymbol{\tau}(\textbf{s})$ is the stress tensor and $\textbf{s}=(\nabla\textbf{u}+\nabla\textbf{u}^\mathrm{T})/2$ is the strain rate tensor, which we decompose into the sum of a base laminar component $\textbf{s}_{0}$ and an additional component due to velocity fluctuations, $\textbf{s}'$. Averaging Eq. \ref{eq1} over time and the entire volume $V$ of the medium then provides a relation for the steady-state pressure drop $
\langle\Delta P\rangle_{t}$ across the medium arising from the sum of these components \cite{SI},
\begin{equation}\label{eq:macrobalance1}
    \frac{\langle\Delta P\rangle_{t}}{\Delta L}\equiv\frac{\eta_{\text{app}} (Q/A)}{k}\approx{\underbrace{\frac{\eta(\dot{\gamma}_{I}) (Q/A)}{k}}
    _{\text{Darcy's law}}}\,
    +{\underbrace{\frac{\langle\chi\rangle_{t,V}}{(Q/A)}}_{\text{Fluctuations}}}.
\end{equation}
    The first term represents Darcy's law for a laminar, steady flow and the second term represents the additional contribution due to unstable flow fluctuations; $\langle\chi\rangle_t\approx\eta\langle\textbf{s}':\textbf{s}'\rangle_t$ quantifies the rate of added viscous dissipation due to unstable flow fluctuations, where $\eta$ is the time-averaged shear viscosity of the polymer solution \cite{SI}. Thus, Eq. \ref{eq:macrobalance1} provides a link between the pore-scale flow fluctuations arising in elastic turbulence and the anomalous increase in macroscopic flow resistance.

\begin{figure}
    \centering
    \includegraphics[width=\textwidth]{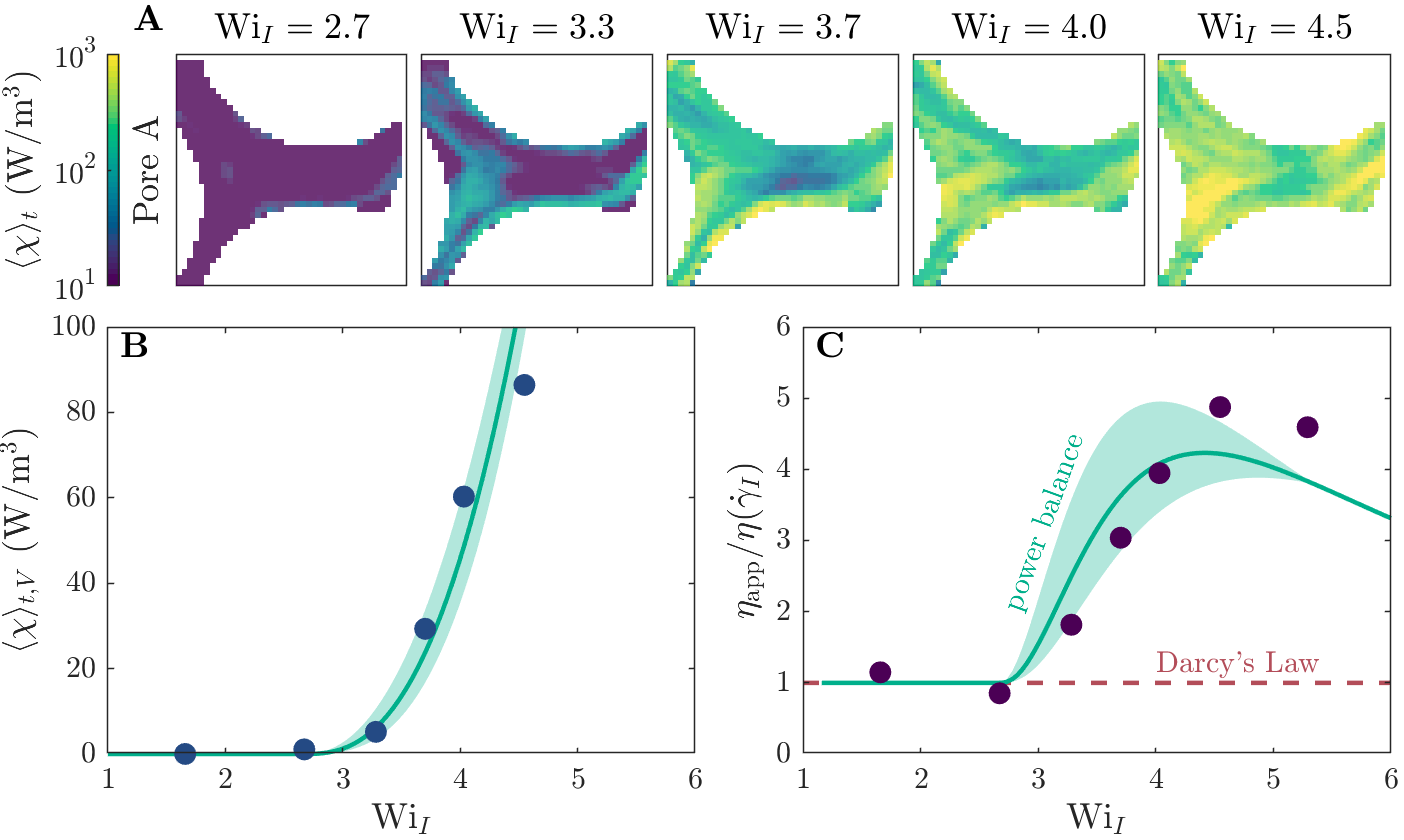}
    \caption{\textbf{Anomalous increase in macroscopic flow resistance is determined by the added viscous dissipation due to unstable flow fluctuations}. \textbf{(A)} The rate of added viscous dissipation $\langle\chi\rangle_t$ directly measured from flow visualization for the example of Pore A sharply increases above the onset of elastic turbulence. \textbf{(B)} Averaging the spatially-averaged $\langle\chi\rangle_t$ over all pores imaged yields the overall added viscous dissipation, which increases as a power law above the macroscopic threshold $\mathrm{Wi}_{c}\approx2.7$. \textbf{(C)} The measured power-law fit to $\langle\chi\rangle_{t,V}$ enables prediction of the macroscopic apparent viscosity $\eta_\mathrm{app}$, as shown by the green curve; the uncertainty associated with the fit to $\langle\chi\rangle_{t,V}$ (shaded region in \textbf{(B)}) yields an uncertainty in this prediction, as shown by the shaded region. Points indicate the independent measurements of $\eta_\mathrm{app}$ from the macroscopic pressure drop.}
    \label{fig:TKE}
\end{figure}

Importantly, the rate of added dissipation $\langle\chi\rangle_t$ can be directly measured using flow visualization in each pore; an example is shown in Fig. 4A. Consistent with our expectation, the rate of local viscous dissipation sharply increases by nearly three orders of magnitude at the onset of elastic turbulence, and continues to increase as $\mathrm{Wi}_{I}$ increases above $\mathrm{Wi}_{c}$ (Fig. 4A, green/yellow regions). These measurements, performed for each pore, thereby enable us to directly test the validity of Eq. \ref{eq:macrobalance1}, assuming spatially isotropic fluctuations within each pore. In particular, we directly compute $\langle\chi\rangle_{t,V}$ by averaging $\langle\chi\rangle_{t}$ over the imaged area of each pore, and then averaging over all the imaged pores. As anticipated, the overall rate of added dissipation increases as a greater fraction of pores becomes unstable (Fig. 4B, symbols), consistent with the power-law scaling shown by the green curve. Incorporating this empirical relationship in Eq. \ref{eq:macrobalance1} then yields a final prediction for the dependence of the apparent viscosity $\eta_{\text{app}}$ on the imposed $\mathrm{Wi}_{I}$ (Fig. 4C, green curve) that is derived directly from our pore-scale imaging of the unstable flow fluctuations. Remarkably, this prediction shows excellent agreement with the macroscopic pressure drop measurements (Fig. 4C, symbols) without employing any fitting parameters. This agreement confirms that the anomalous increase in the macroscopic flow resistance is indeed due to the added dissipation arising from flow fluctuations generated by pore-scale elastic turbulence.

A simple picture for the sigmoidal variation of $\eta_{\text{app}}$ with flow rate, observed in our experiments (Fig. 1B) as well as in numerous previous studies \cite{james1975laminar,durst1981,clarke2016,marshall1967flow}, thereby emerges. At low flow rates, corresponding to $\mathrm{Wi}_{I}<\mathrm{Wi}_{c,min}$, all of the pores in the medium are laminar and steady over time; thus, $\eta_{\text{app}}=\eta(\dot{\gamma}_{I})$. As flow rate is increased, $\mathrm{Wi}_{I}$ eventually exceeds $\mathrm{Wi}_{c,min}\approx2.7$ in our experiments, causing an increasing fraction of pores to become unstable. The added viscous dissipation due to the flow fluctuations in these pores then causes $\eta_{\text{app}}$ to increasingly exceed $\eta(\dot{\gamma}_{I})$ (Fig. 4C, 3\textsuperscript{rd}--6\textsuperscript{th} points). Eventually, as $\mathrm{Wi}_{I}$ exceeds $\mathrm{Wi}_{c,max}\approx4.5$ in our experiments, all of the pores are unstable. Further increases in $\mathrm{Wi}_{I}$ do not appreciably generate additional flow fluctuations, likely due to the finite extensibility of the polymer chains, and $\eta_{\text{app}}$ saturates (Fig. 4C, last point). The steepness of the increase of $\eta_{\text{app}}$ with flow rate therefore reflects the distribution of the different $\mathrm{Wi}_{c}$; while these values depend on the complex 3D geometry of each pore and are challenging to predict \textit{a priori} \cite{pakdel1996}, reducing the polydispersity of the medium likely sharpens the distribution of $\mathrm{Wi}_{c}$ and thus steepens the increase in $\eta_{\text{app}}$, consistent with the results of studies in 2D obstacle arrays \cite{howe2015}. As the flow rate is increased further, we expect that $\eta_{\text{app}}$ eventually converges back to  $\eta(\dot{\gamma})$, reflecting the increased relative influence of viscous dissipation from the base laminar flow---although inertia and chain scission will likely also play a role in this regime, imparting new complexities to the flow. 

Our experiments provide the first visualization of elastic turbulence in 3D porous media. By directly linking pore-scale elastic turbulence to macroscopic transport, our work provides a resolution to the long-standing puzzle of why polymer solutions exhibit an anomalous increase in macroscopic flow resistance in porous media. Our findings that the pore-scale transition to elastic turbulence is a non-equilibrium phase transition akin to the transition to inertial turbulence, and that the resulting dissipation similarly controls macroscopic transport behavior, highlight the connections between these distinct forms of turbulence. Furthermore, by deepening our fundamental understanding of how macroscopic transport behavior depends on imposed flow conditions and solution properties, our analysis provides generally-applicable guidelines for predicting and controlling polymer solution flows in porous media. Because such flows play key roles in removing trapped non-aqueous liquids from subsurface formations during groundwater remediation \cite{smith2008} and oil recovery \cite{sorbie2013,durst1981}, determining separation performance in filtration \cite{bourgeat2003filtration} and chromatography \cite{luo1996high}, improving heat and mass transfer in microfluidic devices \cite{burghelea2004chaotic,ligrani2018heat}, and enabling extrusion-based manufacturing \cite{wang20173d}, we expect these results to impact a broad range of applications.

\newpage
\renewcommand{\thesection}{\Alph{section}}
\renewcommand\thefigure{\thesection\arabic{figure}}
\setcounter{section}{19} 
\setcounter{figure}{0}
\pagenumbering{gobble}

\section*{Materials and Methods}

\subsection{Experimental details}
\subsubsection{Porous medium fabrication and physical characteristics} Our porous medium is a granular packing of borosilicate glass beads with diameters $D_p$ ranging from 300 to 355 $\muup$m (\textit{Mo-Sci}). We pack these grains into a  quartz capillary with a square cross section of area $A=3.1~\text{mm}\times3.1~\text{mm}$ (\textit{Vitrocom}), tap them for a minute to densify, and then lightly sinter the medium in a furnace at 1000\degree C for 3 min. Additionally, we shave down the ends of the packing to provide flat inlets and outlets. This protocol forms a rigid, consolidated, disordered granular packing with a porosity $\phi\approx0.41$, pore throat diameter $d_{t}\approx0.16D_p\approx52~\muup\text{m}$, and tortuosity $\tau\approx2$, as we previously measured using confocal microscopy \cite{datta2013a,krummel2013visualizing}. The length of the medium along the imposed flow direction is $\Delta L=8.1~\mathrm{cm}$. To control and characterize flow in the pore space, we glue inlet and outlet tubing into the inlet and outlet of the medium, respectively, with valves for pressure taps. We determine the medium permeability $k=67~\mu$m$^2$ using Darcy's law, using the value of the pressure drop measured at the lowest (laminar) flow rates; this permeability is in good agreement with our previous measurements of similar porous media \cite{krummel2013visualizing} and with the prediction of the established Kozeny-Carman relation \cite{philipse1993liquid}. 

\subsubsection{Polymer solution preparation and characterization}
Our polymer solution is made by dissolving 18 MDa partially-hydrolyzed polyacrylamide (HPAM; 30\% carboxylated monomers, \textit{Polysciences}) and NaCl (\textit{Sigma Aldrich}) in ultrapure \textit{milliPore} water, and then diluting with glycerol (\textit{Sigma Aldrich}) and dimethyl sulfoxide (DMSO; \textit{Sigma Aldrich}) to obtain a solution whose refractive index is precisely matched to that of the glass beads. The final solution of 300 ppm HPAM, 81 wt\% glycerol, 12 wt\% DMSO, 6 wt\% water, and 1 wt\% NaCl has a measured refractive index of 1.479. All solutions are used within one month of preparation.

We use dynamic light scattering (DLS) to directly measure the mean polymer radius of gyration $R_g$ for our 300 ppm HPAM solution in the same index-matched solvent used in the flow experiments. Four replicate DLS measurements yield $R_g=50\pm30$ nm, indicating that the polymer is polydisperse, likely giving a spectrum of rheological properties and relaxation times. We estimate the polymer overlap concentration \textit{via} the relation $C^{*}\approx (M_{w}/V)/N_{A}$, where $M_{w}$ is the polymer molecular weight, $V=4\pi R_{g}^{3}/3$ is the volume occupied by a single polymer molecule, and $N_{A}$ is Avogadro's number \cite{rubinstein2003polymer}. This calculation yields an overlap concentration of $C^{*}\approx7\%$ and therefore our experiments use a dilute polymer solution at $\approx0.007$ the overlap concentration.

\begin{figure}
    \centering
    \includegraphics[width=\textwidth]{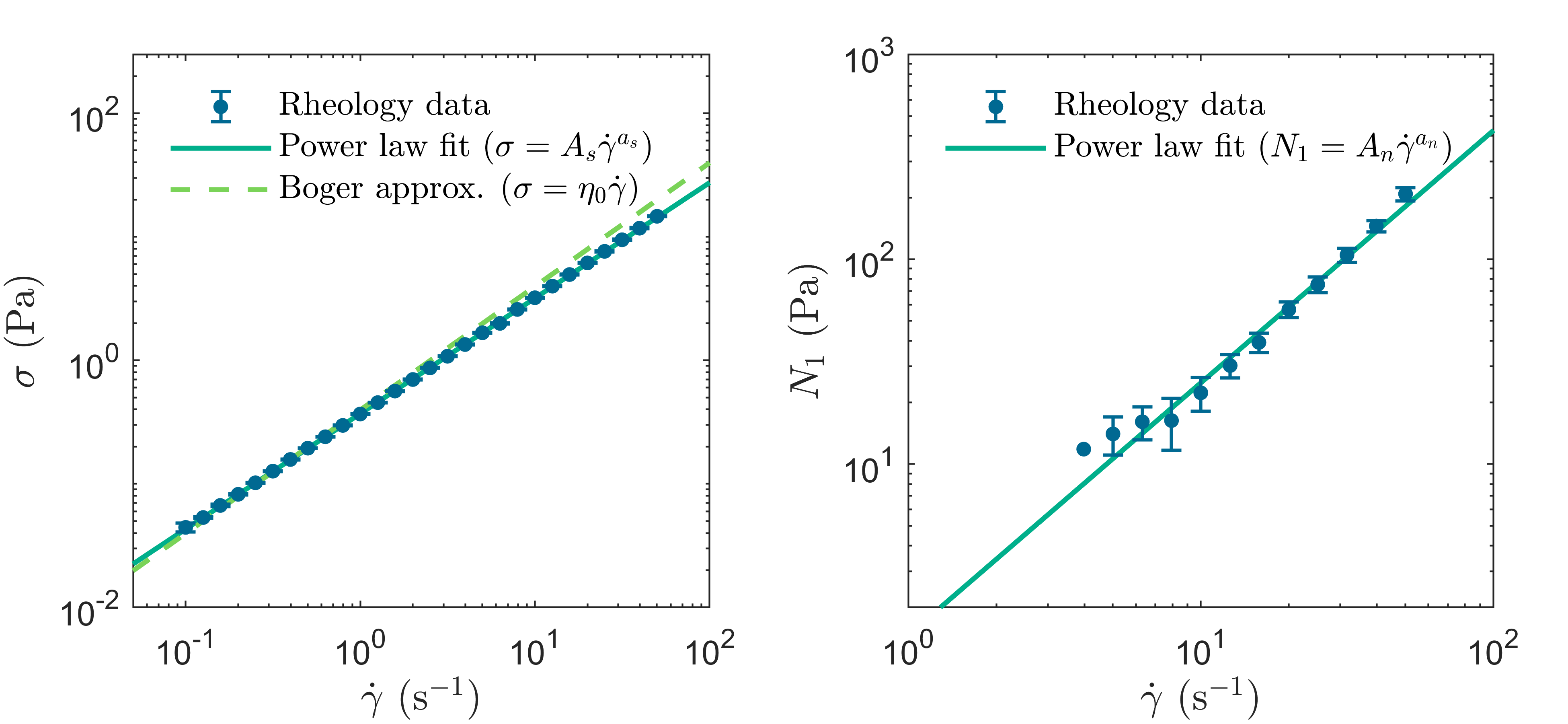}
    \caption{Rheologically measured shear stress and first normal stress difference for the polymer solution used in all experiments. Error bars represent standard deviation over four samples. A power law fit for shear stress $\sigma(\dot{\gamma})\approx A_s(\dot{\gamma})^{\alpha_s}$ gives $A_s\approx0.369(8)$ Pa, $\alpha_s\approx0.934(7)\pm0.001$. A power law fit for the first normal stress difference $N_1(\dot{\gamma})\approx A_n(\dot{\gamma})^{\alpha_n}$ gives $A_n\approx1.46(3)$ Pa, $\alpha_n\approx1.23(1)\pm0.04$.
    }
    \label{fig:SIrheology}
\end{figure}

We characterize all flow properties using shear rheology measurements of a 1 mL sample of the polymer solution. We use a cone-plate geometry in an \textit{Anton Paar} MCR301 rheometer, using a 1\degree~5 cm diameter cone set at a 50 $\muup$m gap. We measure the shear stress $\sigma$ and first normal stress difference $N_{1}$ over a range of shear rates $\dot{\gamma}=0.1~\mathrm{s}^{-1}$ to $50.1~\mathrm{s}^{-1}$, which spans the range of characteristic interstitial shear rates encountered during the flow experiments in porous media, calculated as $\dot{\gamma}_I\equiv Q/\left(A\sqrt{\phi k}\right)$, where $Q$ is the volumetric flow rate of polymer solution, $A$ is the capillary cross sectional area, $\phi$ is the medium porosity, and $k$ is the absolute permeability of the medium; $\sqrt{\phi k}$ therefore defines a characteristic pore dimension. To assess reproducibility, we collect data from four different samples, and find identical results for all four samples. Both the shear stress and first normal stress difference vary with shear rate according to power laws, as shown in Fig. \ref{fig:SIrheology}: $    \sigma(\dot{\gamma})\approx A_s(\dot{\gamma})^{\alpha_s}$ and $N_1(\dot{\gamma})\approx A_n(\dot{\gamma})^{\alpha_n}$, where $\sigma$ and $N_{1}$ have units of Pa, $\dot{\gamma}$ has units of $\mathrm{s}^{-1}$, $A_s=0.369$, $\alpha_s=0.934\pm0.001$,  $A_n=1.46$, $\alpha_n=1.23\pm0.04$. The shear stress varies approximately linearly with shear rate, indicating that shear thinning effects are small due to the high viscosity of the background solvent---which is approximately $\beta=0.2$ times the measured solution viscosity. However, for accuracy, we use the rate-dependent shear viscosity $\eta(\dot{\gamma})\equiv\sigma(\dot{\gamma})/\dot{\gamma}$ in all calculations. We define the zero shear viscosity using the lowest tested interstitial shear rate, $\eta_0\equiv[\sigma(\dot{\gamma})/\dot{\gamma}]_{\dot{\gamma}_I=0.14~\mathrm{s}^{-1}}=0.385~\mathrm{Pa\cdot s}$. 

We define the Reynolds number comparing inertial to viscous stresses as $\mathrm{Re}\equiv\rho (Q/\phi A) d_{t}/\eta(\dot{\gamma})$, where $\rho$ is the density of the solvent. In our porous media experiments, $\mathrm{Re}$ ranges from $2.5\times10^{-6}$ to $1.6\times10^{-4}$    , indicating
that viscous stresses dominate over inertial stresses.

We describe the influence of elasticity using the Weissenberg number, which
compares elastic stresses to viscous stresses. As is conventionally done, we define this parameter as $\mathrm{Wi}\equiv N_1(\dot{\gamma})/2\sigma(\dot{\gamma})$. In our porous media experiments, Wi is greater than one, ranging
from 1.7 to 5.3, indicating that elastic stresses dominate. Moreover, the corresponding
values of the elasticity number $\mathrm{El}\equiv\mathrm{Wi}/\mathrm{Re}$, which compares elastic stresses to inertial
stresses, range from $3.3\times10^{4}$ to $6.7\times10^{5}$, much greater than one. Our experiments thus probe the elasticity-dominated flow regime. Using the shear rheology measurements, we also calculate the rheological relaxation time $\lambda(\dot{\gamma})=\frac{Wi}{\dot{\gamma}}=\frac{N_1(\dot{\gamma})}{2\sigma(\dot{\gamma})\dot{\gamma}}$, whose value ranges from 0.2 to 3 s, in good agreement with previous experiments \cite{qin2017}.

\begin{figure}   
    \centering
    \includegraphics{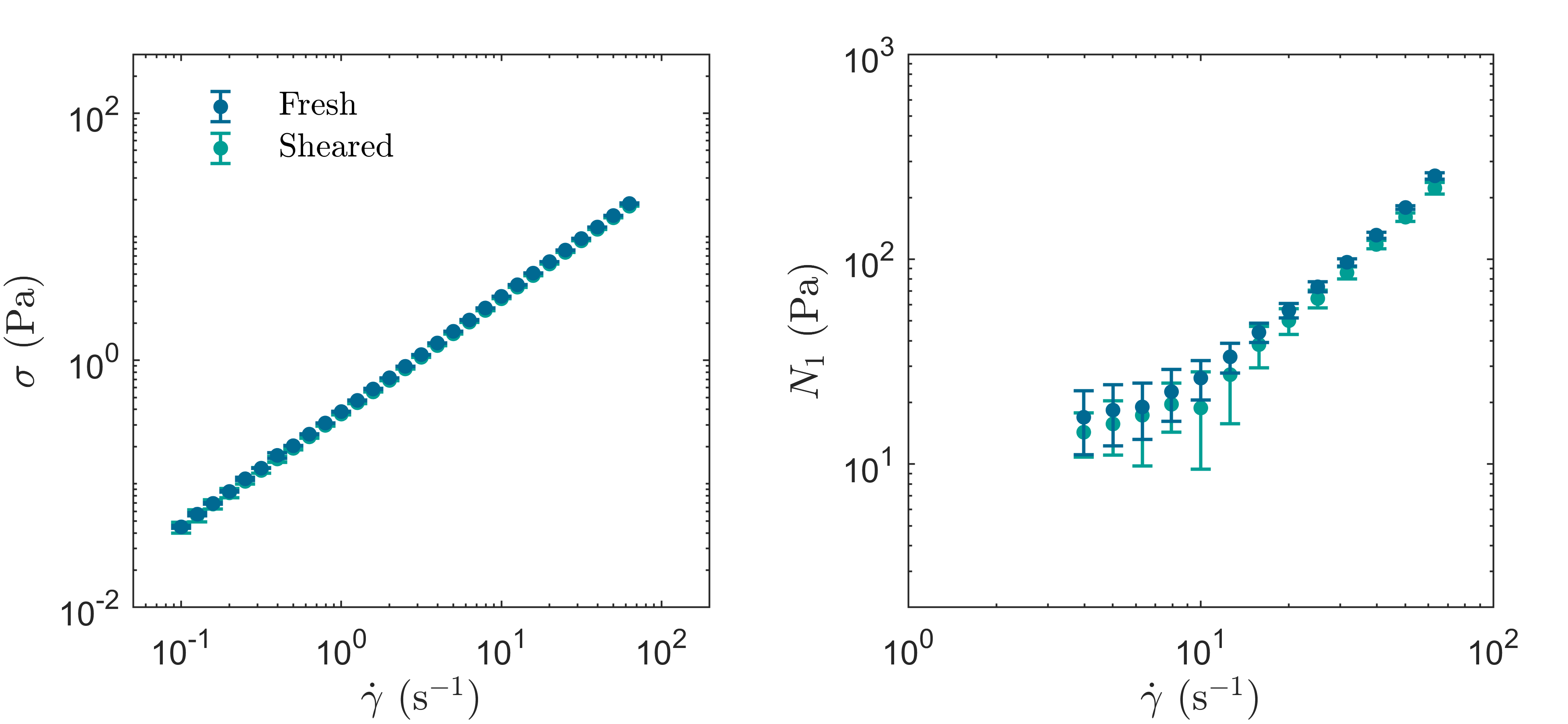}
    \caption{Comparison of rheology for fresh polymer solution and sheared polymer solution passed through porous medium at highest tested flow rate of $Q=5~\mathrm{mL/hr}$. Error bars represent standard deviation over three replicate samples.}
    \label{fig:degRheology}
\end{figure}

To assess possible degradation of polymers due to unstable
flow in the porous media  \cite{vanapalli2006universal}, we also characterize the  rheology of the same polymer solution before and after performing flow experiments at the highest flow rate
tested, $Q=5~\mathrm{mL/hr}$. As shown in Fig. \ref{fig:degRheology}, we do not find observable variation in the shear rheology,  indicating that polymer degradation due to the unstable flow is minimal.

\subsubsection{Characterization of flow in the porous medium}
Before each experiment, we remove air bubbles under vacuum and then fill the medium with water. We then displace the water with the miscible polymer solution, injected into the medium at a constant flow rate $Q$ using a \textit{Harvard Apparatus PhD 2000} syringe pump, for at least three hours to equilibrate the solution in the medium before flow characterization. After each subsequent change in flow rate, the flow is given one hour to equilibrate before characterization. This protocol enables any possible surface adsorption processes to reach a dynamic steady state \cite{parsa2020origin}.

We measure the pressure drop across the medium $\Delta P$ using an \textit{Omega PX26} differential pressure transducer, averaging measurements obtained over 60 min; the fluctuations in these measurements are minimal, and as a result, the corresponding error bars in Fig. 1B of the main text are smaller than the symbol size.

To visualize the pore-scale flow \textit{in situ}, we seed the polymer solution with 5 ppm of fluorescent carboxylated polystyrene tracer particles (\textit{Invitrogen}), $D_t= $ \SI{200}{\nano\meter} in diameter, over 200 times smaller than the pore throat diameter; the particles have excitation between 480 and 510 nm with an excitation peak at 505 nm, and emission between 505 and 540 nm with an emission peak at 515 nm. Particles are tracked using a 488 nm excitation laser, and detected with a 500-550 nm sensor. To visualize the pore space, we also dye the solution with 0.5 ppm of rhodamine red dye, which has an excitation wavelength between 480 and 600 nm with an excitation peak at 560 nm, and emission between 550 and 700 nm with an emission peak at 580 nm. The dyed pore space is imaged using a 561 nm excitation laser, and detected with a 570-620 nm sensor.  Choice of these fluorescent markers allows us to image both the pore space and the dynamic flow within it at high resolution, with no observable cross talk or bleed through on the laser channels. The tracer particles can be considered faithful tracers of the streamlines because their advection dominates over diffusion, as described by the particle-scale P\'eclet number $\mathrm{Pe}\equiv (Q/A)D_p/\mathcal{D}>10^5\gg1$, where $\mathcal{D}=k_BT/3\pi\eta_0D_t=6\times10^{-3}~\muup\mathrm{m}^2/s$ is the Stokes-Einstein particle diffusivity.

We monitor the flow in individual pores using a \textit{Nikon} A1R+ laser scanning confocal fluorescence microscope. We use a 10x objective interrogating a 318 $\muup$m by 318 $\muup$m field of view with the confocal resonant scanner at a temporal resolution of 30 fps and a spatial resolution of 0.62 $\muup$m from an optical slice of $8~\muup\mathrm{m}$ thickness at a depth of $\sim200~\muup\mathrm{m}$ within the medium. To monitor the slow changes in the flow field over time, we record the flow in 2 s intervals every 4 min for 60 min. We repeat this measurement for nine pores randomly chosen near the inlet of the medium and ten pores randomly chosen near the outlet; we do not observe noticeable differences in the results obtained depending on position along the medium. We then measure the two-dimensional velocity field within each pore, with spatial discretization $\Delta x=7.74~\muup\mathrm{m}$, using particle image velocimetry (PIV) for each frame \cite{thielicke2014pivlab}. We observe minimal fluctuations over the course of each 2 s interval, so we average the velocity field obtained in each such interval to give a pseudo-steady snapshot of the velocity field at each time point separated by 4 min, $\mathbf{u}(\mathbf{x},t)\equiv\left(u(\mathbf{x},t),v(\mathbf{x},t)\right)$ where the position vector $\mathbf{x}\equiv(x,y)$. Of the 19 pores imaged, 12 exhibit a well-defined critical $\mathrm{Wi}_c$ below which the pore-scale flow is stable and laminar, above which the pore-scale flow is unstable. We therefore restrict the data shown in Fig. 2 to these 12 pores, which represent $\approx63\%$ of the porous medium. The other 7 pores have $\mathrm{Wi}_c$ above or below the range of Wi explored in our experiments; thus, thus, the distribution shown in Fig. 2F does not account for $\approx37\%$ of the pores of the medium.


\subsection{Analysis of pore-scale flow}

\subsubsection{Processing of PIV data}
In our analysis of the pore-scale flow, the root mean square velocity of a pixel is calculated as the temporal root mean square of the magnitude of the fluctuation from the temporal mean \cite{ruth2019bubble}, $u_\mathrm{rms}(\mathbf{x})=\big(\langle ||\mathbf{u}(\mathbf{x},t)-\langle\mathbf{u}(\mathbf{x},t)\rangle_t||^2 \rangle_t \big)^{1/2}$. We normalize this quantity by the velocity magnitude averaged over time and space (over all pixels) for each pore, $\langle u\rangle_{t,\mathbf{x}}=\langle\langle||\mathbf{u}(\mathbf{x},t)||\rangle_t\rangle_\mathbf{x}$. To quantify velocity fluctuations arising from unstable flow, we compute the velocity fluctuations $\mathbf{u}'(\mathbf{x},t)=\mathbf{u}(\mathbf{x},t)-\langle\mathbf{u}(\mathbf{x},t)\rangle_t$. This fluctuation field enables us to calculate the strain rate tensor associated with flow fluctuations, $s_{ij}'=\partial u'_i/\partial x_j$, pixel-by-pixel. In general, to compute the discrete derivatives, we use the central difference method, in which the derivative of $f$ with respect to $x$ evaluated around $x=x_{0}$ is given by
\begin{align*}
    \bigg(\frac{\partial f}{\partial x}\bigg)_{x_{0}}\approx\frac{1}{2}\bigg(\frac{f(x_0+\Delta x)-f(x_0) }{\Delta x}+\frac{f(x_0)-f(x_0-\Delta x) }{\Delta x}\bigg)
\end{align*}
On the boundaries of data sets, this central difference is replaced with the forward or backward finite difference (first or second term respectively).

\subsubsection{Distributions of key flow parameters} \label{SI:lam}
\begin{figure}
    \centering
    \includegraphics[width=\textwidth]{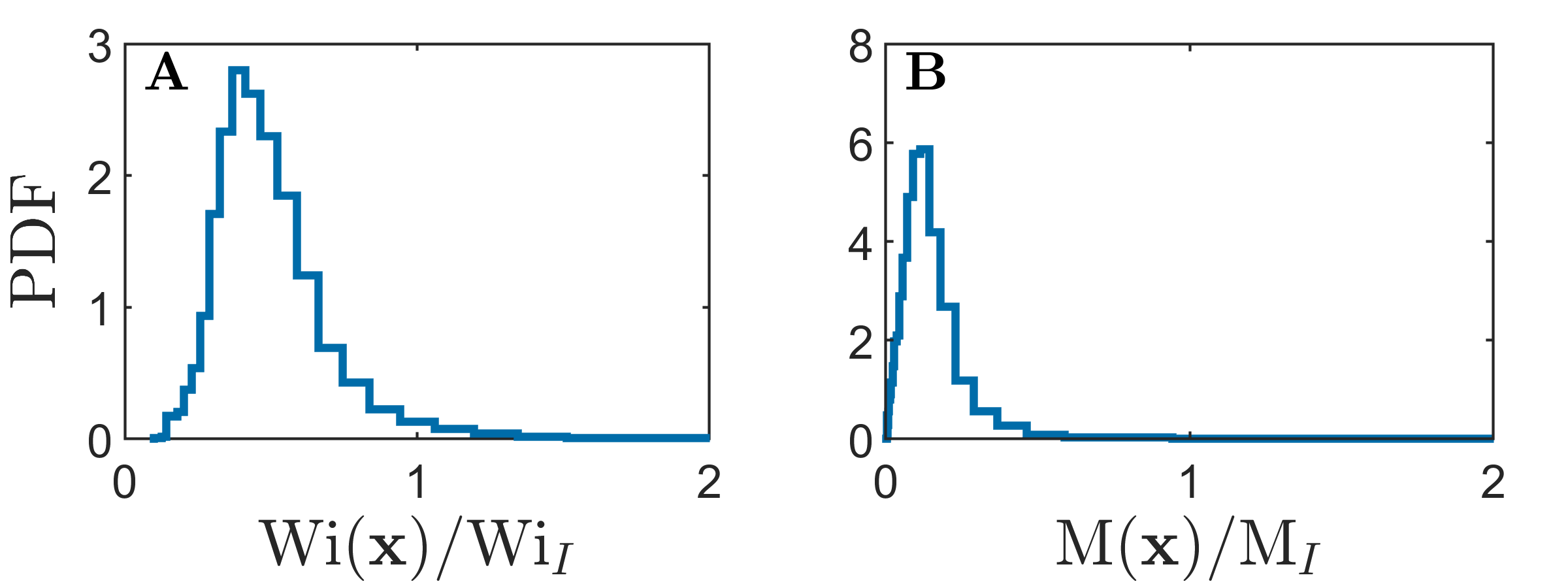}
    \caption{Distributions of flow parameters for 19 imaged pores in the laminar steady flow regime. \textbf{(A)} The local $\mathrm{Wi}$ is broadly distributed; the characteristic macroscopically-defined  $\mathrm{Wi}_I$ represents the upper bound of this distribution. \textbf{(B)} The local $M$ is also broadly distributed; the characteristic macroscopically-defined  $M_I$ represents the upper bound of this distribution.}
    \label{fig:SIdist}
\end{figure}

To characterize the distribution of key flow parameters in the porous medium in the stable laminar case, we use our PIV measurements well below the onset of elastic turbulence (at {$\dot{\gamma}_{I}=0.55~\mathrm{s^{-1}}$}) to determine the base laminar flow field throughout the pore space. We then compute the shear rate $\dot{\gamma}=\partial u/\partial y+\partial v/\partial x$, and use this to compute the spatially-varying Weissenberg number $\mathrm{Wi}(\mathbf{x})\equiv N_1\big(\dot{\gamma}(\mathbf{x})\big)/2\sigma\big(\dot{\gamma}(\mathbf{x})\big)$, pixel-by-pixel, using the rheologically-measured $N_{1}$ and $\sigma$. The distribution of the measured $\mathrm{Wi}(\mathbf{x})$ for 19 imaged pores is shown in Fig. \ref{fig:SIdist}A. As shown by the data, the characteristic interstitial Weissenberg number $\mathrm{Wi}_I\equiv N_1(\dot{\gamma}_I)/2\sigma(\dot{\gamma}_I)$ defined using imposed macroscopic flow conditions and macroscopic characteristics of the porous medium represents the upper limit of this distribution.

Elastic instabilities have been studied in a range of simplified geometries, and are typically parametrized using the Weissenberg number \cite{larson1990,shaqfeh1996,mckinley1996,pakdel1996, rodd2007,rodd2005,teclemariam2007,zilz2012,galindo2012,ribeiro2014,lanzaro2011,lanzaro2014,lanzaro2015,lanzaro2017,qin2019flow,qin2019upstream,qin2017,pan2013,kawale2017a,kawale2017b,varshney2017,kenney2013,shi2015,shi2016,afonso2010purely,sousa2018purely,Browne2020,browne2019pore,groisman2000,groisman2001,groisman2004}. Thus, we also parametrize the different flow rates tested primarily using the Weissenberg number; however, we note that the onset of unstable flow due to streamline curvature can be described using a linear stability analysis of the Stokes equation for a viscoelastic fluid \cite{mckinley1996,pakdel1996}. This analysis indicates that the largest destabilizing term, which leads to the generation of unstable flow locally, is proportional to $M\equiv\sqrt{\mathrm{Wi}\cdot\mathrm{De}}$, where the Deborah number $\mathrm{De}\equiv\lambda(\dot{\gamma}) ||\mathbf{u}||\kappa$ compares the polymer relaxation time $\lambda$ to the flow time scale $\left(||\mathbf{u}||\kappa\right)^{-1}$ and $\kappa$ is a measure of the local streamline curvature \cite{qin2019flow}. In this picture, elastic stresses build up in the flow, generating elastic turbulence when $M$ exceeds a critical value $M_c$, found to be $\approx6$ to $20$ in experiments performed in diverse simplified geometries \cite{pakdel1996,mckinley1996,zilz2012,haward2016elastic,qin2019flow,Browne2020,byars1996experimental}. Thus, the transition to elastic turbulence could also be parameterized using a characteristic interstitial $M_I\equiv\sqrt{\frac{N_1(\dot{\gamma}_I)}{\eta_0\dot{\gamma}_I}\cdot\lambda(\dot{\gamma}_I)(Q/A)\kappa_I}$, again defined using imposed macroscopic flow conditions and macroscopic characteristics of the porous medium; here the characteristic streamline curvature is set by the pore length scale, with $\kappa_I=1/\left(2\sqrt{\phi k}\right)$. We again use our measurements of the spatially-varying shear rate $\dot{\gamma}(\mathbf{x})$, as well as direct measurements of the spatially-varying local streamline curvature $\kappa(\mathbf{x})$, to compute the spatially-varying $M(\mathbf{x})=\sqrt{\mathrm{Wi}\left(\mathbf{x}\right)\cdot\mathrm{De}\left(\mathbf{x}\right)}$, pixel-by-pixel. The distribution of the measured $M(\mathbf{x})$ for all 19 imaged pores is shown in Fig. \ref{fig:SIdist}B. As shown by the data, the characteristic interstitial $M_I$ defined using imposed macroscopic flow conditions and macroscopic characteristics of the porous medium represents the upper limit of this distribution. For our experiments, $\mathrm{M}_I$ ranges from 3.3 to 8.1. The range of $\mathrm{M}_{c,i}$ at which pores become unstable in our experiments is measured to be $\approx$ 5.5 to 7.9, in good agreement with the range of $\approx$ 6 to 20 observed for the onset of unstable flow in simplified geometries.

\begin{figure}
    \centering
    \includegraphics[width=0.6\textwidth]{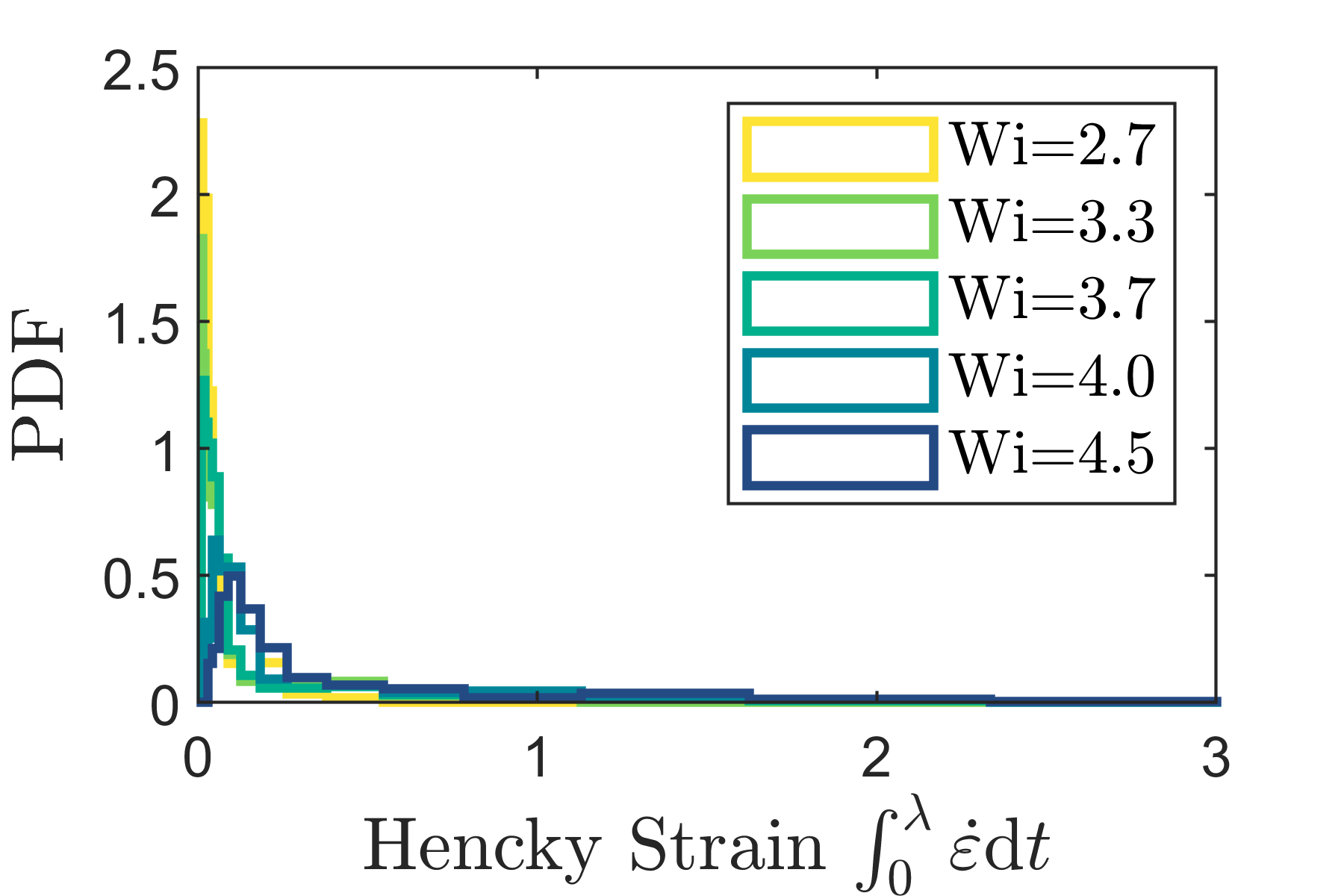}
    \caption{Distribution of observed Hencky strains along sample pathlines of duration $\lambda_0\approx1~\mathrm{s}$. Colors indicate different macroscopic flow rates (reported as $\mathrm{Wi}_I$). Distributions taken over three pores, each with five sample track starting locations, and 15 time points with varying flow fields. }
    \label{fig:henckyDist}
\end{figure}

\subsubsection{Accumulation of Hencky strain}\label{hencky} 
Measurements of a range of polymer solutions indicate that extensional viscosity begins to rise when the accumulated Hencky strain exceeds $\approx2$ to $3$ \cite{mckinley2002filament}. To assess whether extensional viscosity effects are appreciable as polymers navigate the unstable flow in a porous medium, we directly compute the strain history of sample fluid elements along Lagrangian paths. For a selected fluid element voxel, we use the measured time-dependent 2D velocity field to compute its propagation; specifically, using the pixel-by-pixel local velocity $\mathrm{d}\mathbf{x}/\mathrm{d}t=\mathbf{u}(\mathbf{x},t)$, we compute the time to move to the next voxel as $\Delta t\approx\Delta x/u(\mathbf{x},t)$, where $\Delta x$ is the pixel size and $u(\mathbf{x},t)$ is the local velocity magnitude. For this computed pathline, we then compute the accumulated Hencky strain over one polymer relaxation time $\lambda\approx1$ s as $\varepsilon=\int_0^{\lambda}\dot{\varepsilon}(\mathbf{x},t)\mathrm{d}t$. We perform this measurement for five different starting locations near the inlet of a pore, and repeat this set of five measurements for three other pores, to obtain a distribution of Hencky strains. The resulting distributions are shown in Fig. \ref{fig:henckyDist}. Notably, even at the highest $\mathrm{Wi}_I$ tested, $\varepsilon\lesssim1$, suggesting that extensional viscosity effects do not play an appreciable role in the unstable, spatially- and temporally-fluctuating flow: while local polymer extension drives the onset of the unstable flow, accumulated extension is likely not a strong contributor to the global viscous dissipation. Similar behavior is seen in elasto-inertial turbulence, where polymers stretching is highly transient and localized \cite{xi2012intermittent}.

\subsection{Power density balance}
We hypothesize that the flow fluctuations that arise in elastic turbulence impart additional viscous dissipation to the flow—--akin to fluctuations in inertial turbulence \cite{delafosse2011estimation,ha2017estimating}. To quantify this hypothesis, we start with the scalar partial differential equation for the rate of change of mechanical energy per unit volume, obtained by dotting the Cauchy momentum equation with velocity \cite{bird1961transport}:
\begin{align}\tag{S1}
    \frac{\partial}{\partial t}\left(\frac{1}{2}\rho u^2\right) &+\nabla\cdot\frac{1}{2}\rho u^2\mathbf{u}-P\nabla\cdot\mathbf{u}+\nabla\cdot[\boldsymbol{\tau}\cdot\mathbf{u}]-\rho\mathbf{u}\cdot\mathbf{g}=-\nabla\cdot P\mathbf{u}-\boldsymbol{\tau}:\nabla\mathbf{u}
\end{align}
where $\mathbf{u}(\mathbf{x},t)$ is the fluid velocity, $\rho$ is the fluid density, $P(\mathbf{x},t)$ is the fluid pressure, $\boldsymbol{\tau}(\mathbf{x},t)$ is the fluid stress tensor, and $\mathbf{g}$ is gravitational acceleration. 

The first term $\frac{\partial}{\partial t}\left(\frac{1}{2}\rho u^2\right)$ represents the change in kinetic energy, which is of order Re and thus negligible. The second term $\nabla\cdot\frac{1}{2}\rho u^2\mathbf{u}$ represents the acceleration over a control volume; this term disappears, since the inlet and outlet of our capillary have the same surface area, so there is no acceleration across the medium. The third term $P\nabla\cdot\mathbf{u}$ represents the reversible work of compression, which is negligible for an incompressible fluid $\nabla\cdot\mathbf{u}=0$. The fourth term $\nabla\cdot[\boldsymbol{\tau}\cdot\mathbf{u}]$ represents viscous work done across control surfaces; this term disappears, since there is no viscous work done at the capillary walls and the flow is unidirectional across the inlet and outlet control surfaces. The fifth term $\rho\mathbf{u}\cdot\mathbf{g}$ represents gravitational work done, which scales with the Reynolds and Froude numbers: $\rho\mathbf{u}\cdot\mathbf{g}\sim \mathrm{Re}/\mathrm{Fr}^2=\rho g D_p^2/\eta_0\approx0.0028\ll1$ and is thus negligible. This leaves only the last two terms for our experiments: 
\begin{equation}\tag{S2}
    -\nabla\cdot P\mathbf{u}=\boldsymbol{\tau}:\nabla\mathbf{u}
\label{maineq}
\end{equation}
 The left hand side represents the rate of work done by the fluid pressure and the right hand side represents the rate of viscous energy dissipation, per unit volume. The velocity gradient tensor can be decomposed into a symmetric and asymmetric component $\nabla\mathbf{u}=\mathbf{s}+\boldsymbol{\omega}$, where $\mathbf{s}=(\nabla\mathbf{u}+\nabla\mathbf{u}^\mathrm{T})/2$ is the rate of strain tensor and $\boldsymbol{\omega}=(\nabla\mathbf{u}-\nabla\mathbf{u}^\mathrm{T})/2$ is the vorticity tensor.

\subsubsection{Macroscopic averaging}
Taking the volume integral of Eq. \ref{maineq} and applying the divergence theorem to the left hand side yields the macroscopic power balance over the control volume. This volume is composed of the four capillary walls and a surface perpendicular to the walls well upstream and downstream of the bead packing, such that the flow is unidirectional $\mathbf{u}=u_x\hat{\mathbf{x}}$ across the inlet/outlet surfaces $\mathbf{n}=\pm\hat{\mathbf{x}}$:
\begin{align}
    -\int_\mathscr{A} P\mathbf{u}\cdot\mathbf{n}\mathrm{d}A=\int_V\boldsymbol{\tau}:(\mathbf{s}+\boldsymbol{\omega})\mathrm{d}V\nonumber\nonumber\\
    \implies(Q/A)A{\Delta P} =V\langle\boldsymbol{\tau}:(\mathbf{s}+\boldsymbol{\omega})\rangle_V\nonumber\\
    \implies\frac{\Delta P}{\Delta L} =\frac{\langle\boldsymbol{\tau}:(\mathbf{s}+\boldsymbol{\omega})\rangle_V}{Q/A}.\tag{S3}
\label{volumeaveraged}
\end{align}

\subsubsection{Time averaging}
Drawing inspiration from the treatment of inertial turbulence, in which flows similarly exhibit strong spatio-temporal fluctuations, we decompose the velocity into a time-averaged and a fluctuating component $\mathbf{u}(\mathbf{x},t)=\mathbf{u}_0(\mathbf{x})+\mathbf{u}'(\mathbf{x},t)$, from which it follows that the rate of strain and vorticity tensors also decompose $\mathbf{s}(\mathbf{x},t)=\mathbf{s}_0(\mathbf{x})+\mathbf{s}'(\mathbf{x},t)$ and $\boldsymbol{\omega}(\mathbf{x},t)=\boldsymbol{\omega}_0(\mathbf{x})+\boldsymbol{\omega}'(\mathbf{x},t)$. The pressure similarly decomposes into a mean and fluctuating component $P(\mathbf{x},t)=P_0(\mathbf{x})+P'(\mathbf{x},t)$, with $\langle P'\rangle_t=0$ and thus $\langle P\rangle_t=P_{0}$. Time averages represent the time average $\langle~~\rangle_t=\frac{1}{t_c}\int^{+t_c/2}_{-t_c/2}(~~)\mathrm{d}t$ of Eq. \ref{volumeaveraged} over a moving window $t=\pm t_c/2$, where $t_c$ is a sufficiently large time window for meaningful averaging  \cite{whitaker1992introduction}:

\begin{equation}\label{eq:macrobalance2}\tag{S4}
    \frac{\langle\Delta P\rangle_{t}}{\Delta L} =\frac{\langle\langle{\boldsymbol{\tau}:(\mathbf{s}+\boldsymbol{\omega})}\rangle_{t}\rangle_V}{Q/A}.
\end{equation}

We then decompose the dissipation function $\langle\boldsymbol{\tau}:\nabla\mathbf{u}\rangle_t$ into a mean and fluctuating component. Because our calculations of Hencky strain (described in Section \ref{hencky}) suggest that extensional viscosity does not appreciably contribute to the global viscous dissipation, we express the fluid  stress as a function of the rate of strain tensor, $\tau_{ij}(s_{ij})$ \cite{zami2016transition}. Since the stress is nonlinear for a non-Newtonian fluid, the function for stress $\tau_{ij}(s_{0,ij}+s'_{ij})$ cannot easily be separated into a mean and fluctuating term; instead, we expand $\tau_{ij}$ with a Maclaurin series, applying the definition of fluctuations $\langle s_{ij}'\rangle_t\equiv0$ and $\langle \omega_{ij}'\rangle_t\equiv0$, but $\langle s_{ij}'^2\rangle_t\neq0$:
\begin{align}
    \langle {\tau_{ij}}&|_{{s}_{0,ij}+{s}'_{ij}}({s}_{ij}+{\omega}_{ij})\rangle_t\nonumber\\
    &=\bigg\langle\bigg({\tau}_{ij}|_{{s}_{0,ij}}
    + \frac{\partial{\tau}_{ij}}{\partial{s}_{ij}}\bigg|_{{s}_{0,ij}}{s}'_{ij}    +\frac{1}{2}\frac{\partial^2{\tau}_{ij}}{\partial{s}_{ij}^2}\bigg|_{{s}_{0,ij}}{s}'^2_{ij}
    +\frac{1}{3}\frac{\partial^3{\tau}_{ij}}{\partial{s}^3_{ij}}\bigg|_{{s}_{0,ij}}{s}'^3_{ij}
    +\mathscr{O}({s}'^4_{ij})\bigg)\big({s}_{0,ij}+{\omega}_{0,ij}+{s}'_{ij}+{\omega}'_{ij}\big)\bigg\rangle_t\nonumber\\
    &={\underbrace{{\tau}_{ij}|_{{s}_{0,ij}}\big({s}_{0,ij}+{\omega}_{0,ij}\big)}_{\text{Mean flow: Darcy}}}
    +{\underbrace{\Bigg[\frac{\partial{\tau}_{ij}}{\partial{s}_{ij}}\bigg|_{{s}_{0,ij}}+\frac{{s}_{0,ij}+{\omega}_{0,ij}}{2}\frac{\partial^2{\tau}_{ij}}{\partial{s}^2_{ij}}\bigg|_{{s}_{0,ij}}\Bigg]\langle{{s}'^2_{ij}}\rangle_t}_{\text{Unstable flow:}~ \langle\chi\rangle_t}}
    +{\mathscr{O}(\langle{s_{ij}'^4}\rangle_t)},\tag{S5}
    \label{maclaurin}
\end{align}
which is accurate to fourth order $\mathscr{O}(\langle{s'^4_{ij}}\rangle_t)$. The first term reflects the viscous dissipation of the mean flow, ultimately yielding Darcy's law when volume averaged, by definition: $\langle{\tau}_{ij}|_{{s}_{0,ij}}({s}_{0,ij}+{\omega}_{0,ij})\rangle_V/(Q/A)=\eta(\dot{\gamma}_I)(Q/A)/k$. The second term reflects viscous dissipation due to unstable flow fluctuations, and we define it as the rate of added dissipation $\langle\chi\rangle_t$.

\subsubsection{Unstable dissipation function}\label{SI:chi}
The term in square brackets in Eq. \ref{maclaurin} has units of a dynamic viscosity, prompting the \textit{ansatz} that it should be proportional to $\eta(\dot{\gamma}_0)$, where $\dot{\gamma}_0\equiv 2s_{0,xy}=\partial u_0/\partial y+\partial v_0/\partial x$ and $c_{ij}$ is the proportionality constant:
\begin{align}
    \langle\chi\rangle_t&\equiv\Bigg[\frac{\partial{\tau}_{ij}}{\partial{s}_{ij}}\bigg|_{{s}_{0,ij}}+\frac{{s}_{0,ij}+{\omega}_{0,ij}}{2}\frac{\partial^2{\tau}_{ij}}{\partial{s}^2_{ij}}\bigg|_{{s}_{0,ij}}\Bigg]\langle{{s}'^2_{ij}}\rangle_t\nonumber\\
    &\equiv c_{ij}\eta(\dot{\gamma}_0)\langle{{s}'^2_{ij}}\rangle_t.\tag{S6}
\end{align}
For a power-law fluid, $\tau_{ij}=A_s(s_{ij})^{\alpha_s}$, where $A_s$ and $\alpha_s$ are material constants. This constitutive relationship allows us to compute $c_{ij}$: 
\begin{align}
    c_{ij}\eta(\dot{\gamma}_0)&\equiv\frac{\partial{\tau}_{ij}}{\partial{s}_{ij}}\bigg|_{{s}_{0,ij}}+\frac{{s}_{0,ij}+{\omega}_{0,ij}}{2}\frac{\partial^2{\tau}_{ij}}{\partial{s}^2_{ij}}\bigg|_{{s}_{0,ij}}\nonumber\\
    &=\alpha_sA_s{s}_{0,ij}^{\alpha_s-1}\bigg(1+\frac{{s}_{0,ij}+{\omega}_{0,ij}}{{s}_{0,ij}}\frac{(\alpha_s-1)}{2}\bigg)\nonumber\\
    &={\alpha_s}{2^{1-\alpha_s}}\bigg(1-(1+\Lambda_{ij})\frac{(1-\alpha_s)}{2}\bigg)\eta(\dot{\gamma}_0)\tag{S7}
\end{align}
where, assuming isotropic unstable flow fluctuations, $\eta(s_{0,ij})\approx\eta(s_{0,xy})\equiv\eta(\dot{\gamma}_{0}/2)$.

\begin{figure}
    \centering
    \includegraphics[width=3in]{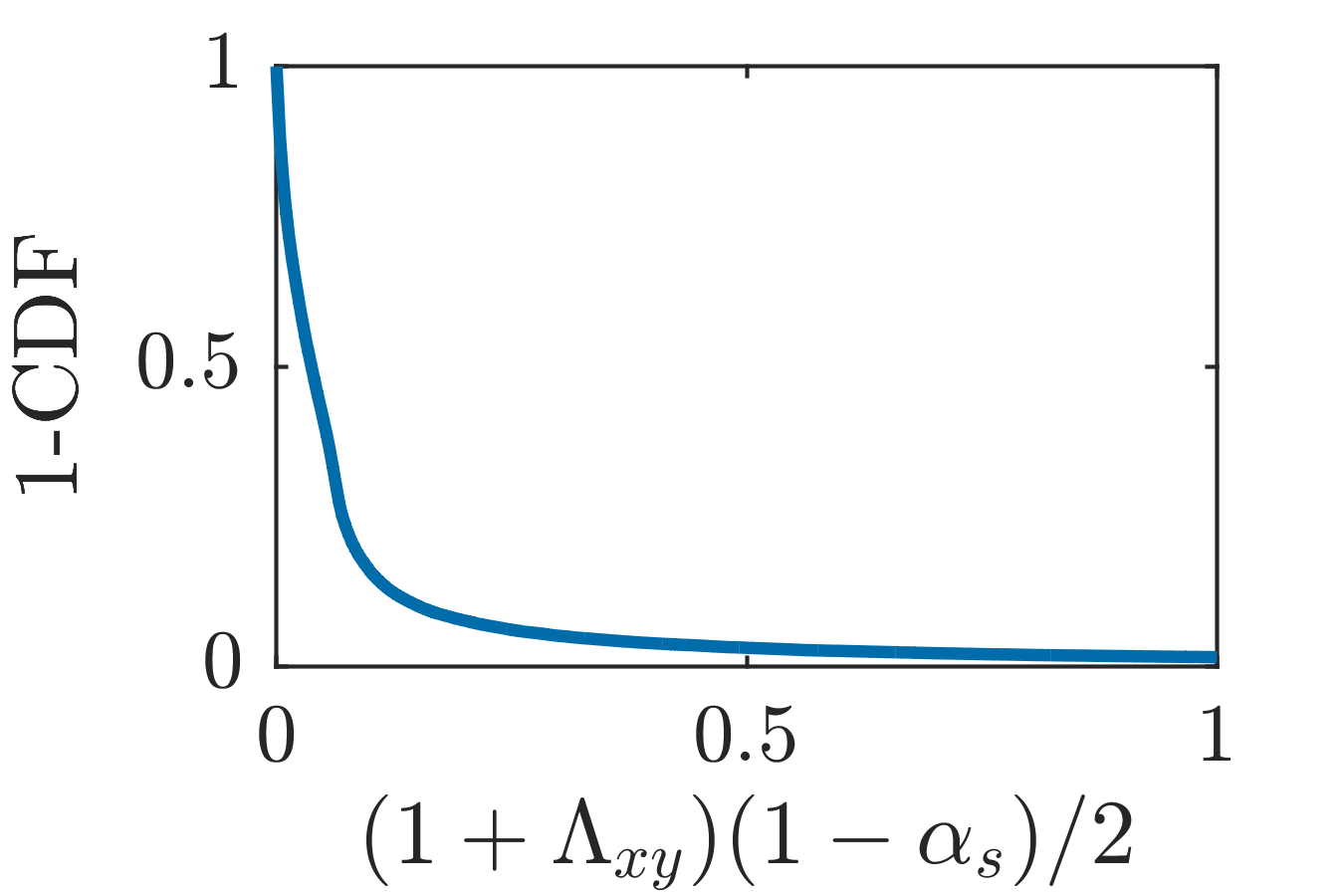}
    \caption{The complementary cumulative distribution function of the in-plane component of the correctional term $(1+\Lambda_{xy})(1-\alpha_s)/2$, distributed over all tested flowrates and pixels. For a vast majority of pixels, the magnitude of this term is much less than 1. The average value of $\langle(1+\Lambda_{xy})(1-\alpha_s)/2\rangle_{v,Q}=0.026\ll1$ indicates that $1-(1+\Lambda_{xy})(1-\alpha_s)/2\approx1$.}
    \label{fig:VortDist}
\end{figure}

The term $\Lambda_{ij}\equiv{\omega}_{0,ij}/{{s}_{0,ij}}$ cannot be directly measured from a 2D flow field; simple averaging for the unknown elements of would trivially return $\Lambda_{ik}=0$. However, estimating the magnitude of $\Lambda_{ij}$ using just the in-plane component indicates that the entire term is typically much less than order one: averaging over all pixels and flow rates yields $\langle(1+\Lambda_{ij})(\alpha_s-1)/2\rangle_{V,Q}=0.026\ll1$, as shown in Fig. \ref{fig:VortDist}. We therefore neglect this term. Thus, $c=\alpha_s2^{1-\alpha_s}$; $c=1$ for a Newtonian fluid and $0<c<1$ for shear-thinning fluids. Using our measured fluid rheology, we find $c=0.98$---reflecting that our fluid has nearly constant shear viscosity for the shear rates tested.

The unstable dissipation function $\langle\chi\rangle_t$ then depends primarily on the fluctuating rate of strain tensor $\langle s'^2_{ij}\rangle_t$. Again assuming isotropic flow fluctuations, as is frequently done in the case of inertial turbulence \cite{delafosse2011estimation,sharp2000dissipation},
\begin{align}
    \bigg\langle\bigg(\frac{\partial u'_z}{\partial z}\bigg)^2\bigg\rangle_t&\approx
    \frac{1}{2}\Bigg[\bigg\langle\bigg(\frac{\partial u'_x}{\partial x}\bigg)^2\bigg\rangle_t+\bigg\langle\bigg(\frac{\partial u'_y}{\partial y}\bigg)^2\bigg\rangle_t\Bigg]\nonumber\\
    \bigg\langle\bigg(\frac{\partial u'_x}{\partial z}\bigg)^2\bigg\rangle_t&\approx\bigg\langle\bigg(\frac{\partial u'_y}{\partial z}\bigg)^2\bigg\rangle_t\approx\bigg\langle\bigg(\frac{\partial u'_z}{\partial x}\bigg)^2\bigg\rangle_t\approx\bigg\langle\bigg(\frac{\partial u'_z}{\partial y}\bigg)^2\bigg\rangle_t\nonumber\\
    &\approx\frac{1}{2}\Bigg[\bigg\langle\bigg(\frac{\partial u'_x}{\partial y}\bigg)^2\bigg\rangle_t+\bigg\langle\bigg(\frac{\partial u'_y}{\partial x}\bigg)^2\bigg\rangle_t\Bigg]\nonumber\\
    \bigg\langle\bigg(\frac{\partial u'_x}{\partial z}\frac{\partial u'_z}{\partial x}\bigg)\bigg\rangle_t&\approx\bigg\langle\bigg(\frac{\partial u'_y}{\partial z}\frac{\partial u'_z}{\partial y}\bigg)\bigg\rangle_t\nonumber\\
    &\approx-\frac{1}{4}\Bigg[\bigg\langle\bigg(\frac{\partial u'_x}{\partial x}\bigg)^2\bigg\rangle_t+\bigg\langle\bigg(\frac{\partial u'_y}{\partial y}\bigg)^2\bigg\rangle_t\Bigg]\nonumber
\end{align}
\begin{align}\label{eq:chi}
    \implies\langle\chi\rangle_t&\equiv c\eta(\dot{\gamma}_0 )\langle{ s'^2_{ij}}\rangle_t\nonumber\\ 
    &\approx c\eta(\dot{\gamma}_0 )\Bigg[2\bigg\langle{\bigg(\frac{\partial u_x'}{\partial x}\bigg)^2}\bigg\rangle_t+2\bigg\langle{\bigg(\frac{\partial u_y'}{\partial y}\bigg)^2}\bigg\rangle_t
    +3\bigg\langle{\bigg(\frac{\partial u_y'}{\partial x}\bigg)^2}\bigg\rangle_t+3\bigg\langle{\bigg(\frac{\partial u_x'}{\partial y}\bigg)^2}\bigg\rangle_t
    +2\bigg\langle{\frac{\partial u_y'}{\partial x}\frac{\partial u_x'}{\partial y}}\bigg\rangle_t\Bigg].\tag{S8}
\end{align}

This quantity, which quantifies the rate of added viscous dissipation due to unstable flow fluctuations, can now be fully determined from our PIV measurements. In the main text, we write this in the form $\langle\chi\rangle_t\approx\eta\langle\textbf{s}':\textbf{s}'\rangle_t$ for simplicity, and our computations use the full form shown in Eq. \ref{eq:chi}.

\subsubsection{Apparent viscosity}
Having computed the unstable dissipation rate $\langle\chi\rangle_t$ using our direct pore-scale flow visualization, \textit{via} Eq. \ref{eq:chi}, we use this quantity to determine the overall apparent viscosity of the flowing polymer solution. First, we directly compute $\langle\chi\rangle_{t,V}$ by averaging $\langle\chi\rangle_{t}$ over the imaged area of each pore, and then averaging over all the imaged pores. Above the critical Weissenberg number $Wi_c=2.7$, $\langle\chi\rangle_{t,V}$ increases sharply with an apparent power law scaling $\langle\chi\rangle_{t,V}=A_x(\mathrm{Wi}/\mathrm{Wi}_c-1)^{\alpha_x}$. We fit $A_x=280\pm1~\mathrm{W/m^3}$ and ${\alpha_x}=2.6\pm0.4$, as shown in Fig. 4B. Then, we substitute $\langle\chi\rangle_{t,V}$ into Eqs. \ref{eq:macrobalance2}-\ref{maclaurin} to obtain our final result:
\begin{align}\label{eq:macrobalance3}
\frac{\langle\Delta P\rangle_{t}}{\Delta L} 
&=\frac{\big\langle\boldsymbol{\tau}|_{\mathbf{s}_0}:\nabla\mathbf{u}_0\big\rangle_V}{Q/A}+\frac{\langle\chi\rangle_{t,V}}{Q/A}\nonumber\\
    &=\frac{\eta(\dot{\gamma}_I)Q/A}{k}+\frac{\langle\chi\rangle_{t,V}}{Q/A}\nonumber\\
    &\implies\eta_{\text{app}}(\dot{\gamma}_I)=\eta(\dot{\gamma}_I)+\frac{k\langle\chi\rangle_{t,V}}{\left(Q/A\right)^2}.\tag{S9}
\end{align}

\newpage\section*{Movie Captions}
\noindent\textbf{S1}: Velocity field of example pore (pore B) just below onset of instability ($\dot{\gamma}_I=2.8~\mathrm{s}^{-1}$; $\mathrm{Wi}_I=2.7$). Applied flow is left to right. Each frame is 4 min apart (720x speed). Arrows indicate the vector field, and colors indicate velocity magnitude as measured by particle image velocimetry (PIV). Velocities do not change appreciably over time above the error of PIV.  
\\
\\
\noindent\textbf{S2}: Velocity field of example pore (pore B) above onset of instability ($\dot{\gamma}_I=8.3~\mathrm{s}^{-1}$; $\mathrm{Wi}_I=3.7$). Applied flow is left to right. Each frame is 4 min apart (720x speed). Arrows indicate the vector field, and colors indicate velocity magnitude as measured by particle image velocimetry (PIV). Velocities exhibit strong spatio-temporal fluctuations, consistent with the onset of an elastic instability. 
\\
\\
\noindent\textbf{S3}: Fluctuating velocity field of example pore (pore B) near cusp of instability ($\dot{\gamma}_I=5.5~\mathrm{s}^{-1}$; $\mathrm{Wi}_I=3.3$). Applied flow is left to right. Each frame is 4 min apart (720x speed). Colors indicate fluctuating velocity magnitude as measured by particle image velocimetry (PIV). Right shows kymograph of fluctuating velocity field for an example column of pixels (marked by red lines). Puffs of fluctuations decay in time.  
\\
\\
\noindent\textbf{S4}: Fluctuating velocity field of example pore (pore B) well above onset of instability ($\dot{\gamma}_I=11~\mathrm{s}^{-1}$; $\mathrm{Wi}_I=4.0$). Applied flow is left to right. Each frame is 4 min apart (720x speed). Colors indicate fluctuating velocity magnitude as measured by particle image velocimetry (PIV). Right shows kymograph of fluctuating velocity field for an example column of pixels (marked by red lines). Fluctuations are sustained in time.
\\
\\
\noindent\textbf{S5}. Fluctuating velocity field of example pore (pore B) well above onset of instability ($\dot{\gamma}_I=6~\mathrm{s}^{-1}$; $\mathrm{Wi}_I=3$) shown at high time resolution. Applied flow is left to right. Each PIV frame averaged over over 1/6 s. Video shown at 5x speed. Colors indicate fluctuating velocity magnitude as measured by particle image velocimetry (PIV). Right shows kymograph of fluctuating velocity field for an example column of pixels (marked by red lines). Fluctuations are sustained in time.


\bibliographystyle{Science}

\newpage\section*{Acknowledgments}
Acknowledgment is made to the Donors of the American Chemical Society Petroleum Research Fund for partial support of this research through grant PRF 59026-DNI9. This material is also based upon work supported by the National Science Foundation Graduate Research Fellowship Program (to C.A.B.) under Grant No. DGE1656466. Any opinions, findings, and conclusions or recommendations expressed in this
material are those of the authors and do not necessarily reflect the views of the National
Science Foundation. C.A.B. was also supported in part by the Mary and Randall Hack
Graduate Award of the High Meadows Environmental Institute.\\


\noindent \textbf{Author contributions:} C.A.B. performed all experiments; C.A.B. and S.S.D. designed the experiments, analyzed the data, developed and implemented the theoretical model, discussed the results, and wrote the manuscript. S.S.D. designed and supervised the overall project.\\

\noindent \textbf{Competing interests:} The authors declare no competing interests.\\

\noindent \textbf{Data and materials availability:} All data are available in the manuscript or the supplementary materials.

\section*{Supplementary Materials}
\noindent Materials and Methods

\noindent Fig S1--S5

\noindent References 37--69

\noindent Movies S1--S5


\end{document}